\documentstyle[epsf,rotate]{mn}

\newcommand{\kms}{\mbox{\,km\,s$^{-1}$}}
\newcommand{\ergs}{\mbox{\,erg\,s$^{-1}$}}
 
\title[Irradiation of the secondary star in X-ray Nova
Scorpii 1994 (=GRO J1655--40)]
{Irradiation of the secondary star in X-ray Nova
Scorpii 1994 (=GRO J1655--40)}
\author[T.~Shahbaz, P.~Groot, S.N.~Phillips, 
J.~Casares, P.A.~Charles, J.~van Paradijs]{
T.~Shahbaz$^{1}$, P.~Groot$^{2}$, S.N.~Phillips$^{1}$, 
J.~Casares$^{3}$, P.A.~Charles$^{1}$, J.~van 
Paradijs$^{2,4}$\\
$^{1}$University of Oxford, Department of Physics, Nuclear Physics
Laboratory, Keble Road, Oxford, OX1 3RH, UK \\
$^{2}$Astronomical Institute ``Anton Pannekoek'', University of
Amsterdam and Center for High Energy Astrophysics, Kruislaan, 403, 1098 SJ \\
Amsterdam, The Netherlands \\
$^{3}$Instituto de Astrof\'\i{}sica de Canarias 38200 La Laguna, 
Tenerife, Spain \\
$^{4}$Physics Department, UAH, Hunstville, Alabama 35899, USA }

\begin{document}

\maketitle

\begin{abstract}

\noindent
We have obtained intermediate resolution optical spectra of
the black-hole candidate Nova Sco 1994 in June 1996, when the source
was in an X-ray/optical active state ($R\sim$15.05).  We measure the
radial velocity curve of the secondary star and obtain a
semi-amplitude of 279$\pm$10\kms;  a value which is 30 per cent larger
than the value obtained when the source  is in quiescence. Our large
value for $K_{2}$ is consistent with 60$^{+9}_{-7}$ per cent of 
the secondary star's
surface being heated; compared to 35 per cent, which is what one would
expect if only the inner face of the secondary star were
irradiated. Effects such as irradiation-induced flows on the
secondary star may be important in explaining the observed large value
for $K_{2}$.

\end{abstract}


\section{Introduction}

The soft X-ray transients, a subclass of the low-mass X-ray binaries
distinguished by their X-ray outbursts, have proved to be an ideal
hunting ground for stellar-mass black hole candidates (Tanaka \&
Shibazaki  1996). The system Nova Sco 1994 (=GRO~J1655--40) is
particularly interesting, since as well as being a source of
superluminal jets (Zhang et al., 1994; Harmon et al., 1995), its
optical brightness and partial eclipse features mean that it is one of
the few systems that has yielded a reliable estimate for the mass of
the collapsed star.

Nova Sco 1994 was discovered on July 27 1994 with BATSE on board the
Compton Gamma Ray Observatory (Zhang et al., 1994). It has been
studied extensively during the past few years in X-rays and at optical
and radio wavelengths (Bailyn et al., 1995a and b, Zhang et al., 1995,
van der Hooft et al. 1998). Strong evidence that the compact object
in Nova Sco 1994 is a black hole was presented by Bailyn et al.\
(1995b) who initially established a spectroscopic period of
$2.601\pm0.027$ days, classified the secondary as an F2--F6$\sc iv$
type star and suggested a mass function $f(M)$=3.16$\pm$0.15
M$_{\odot}$. An improved value of $f(M)$=3.24$\pm$0.09 M$_{\odot}$ was
presented by Orosz \& Bailyn (1997) using both quiescent and outburst
data, derived from a radial velocity semi-amplitude of 228.2$\pm$2.2
\kms. Shahbaz et al. (1999) using only quiescent data,  determined the
true radial velocity semi-amplitude $K_{2}$=215.5$\pm$2.4\kms which
gives a revised value for the mass function of $f(M)$ = 2.73$\pm$0.09
M$_{\odot}$. They also measured the rotational broadening of the
secondary star which then gives the binary mass ratio $q\sim 0.39$
(=M$_{2}$/M$_{1}$, where M$_{1}$ and M$_{2}$ are the masses of the
compact object and secondary star respectively).

The effect of heating of the secondary is to shift the `effective centre'
of the secondary, weighted by the strength of the absorption lines,
from the centre of mass of the star. One expects that this  results in
a significant distortion of the radial velocity curve and renders a
sinusoidal fit clearly inadequate, leading to a spuriously high
radial velocity semi-amplitude. In order to quantify this effect we
have determined the radial velocity variations of the secondary star
in Nova Sco 1994, when it was in outburst and compared our results with
others obtained using data taken when the source was in different
X-ray states.

\section{Observations and Data reduction}

\subsection{Spectroscopy}

Intermediate resolution optical spectra of Nova Sco 1994 were obtained
on 1996 June 20--24 with the 1.54-m Danish Telescope at the European
Southern Observatory (ESO) in Chile using the Danish Faint Object
Spectrograph and Camera (DFOSC). We used grating \#8 which gave a
dispersion of 1.26~\AA\ per pixel and a wavelength coverage 
from 5865--8336\AA.
The Loral 2048$\times$2048 CCD was
used, binned by a factor two in the spatial direction in order to
reduce the readout noise, but not binned in  the dispersion direction.
The seeing during the observations was poor and variable (see section
2.2) so we used a slit width of $2\farcs5$ on the first night and
then  $2\farcs0$ for the other nights. This resulted in spectral
resolutions  of 7.6~\AA\ and 5.5~\AA\ for the first and other nights
respectively. Wavelength calibration was performed using a Cu-Ar arc. 
A total of 47 spectra were taken each having exposure
times of 1800s (see Table 1 for details).

The data reduction and analysis was performed using the Starlink {\sc
figaro} package, the {\sc pamela} routines of K.\,Horne and the {\sc
molly} package of T.\,R.\ Marsh. Removal of the individual bias signal
was achieved through subtraction of a median bias frame. Small scale
pixel-to-pixel sensitivity variations were removed with a flat-field
frame prepared from observations of a tungsten lamp. One-dimensional
spectra were extracted using the optimal-extraction algorithm of Horne
(1986), and calibration of the wavelength scale was achieved using 5th
order polynomial fits which gave an rms scatter of 0.03~\AA. The
stability of the final calibration was verified with the OH sky line
at 6300.3\AA\ whose position was accurate to within 0.1 \AA.

\subsection{Photometry}

Using the same setup as for the spectroscopy, we also obtained
limited Bessell $r$-band images of Nova Sco 1994 every night.  The
data were debiased using a median bias frame, but not flat-fielded, as
none were taken.  These images were used to estimate the seeing each
night (see Table 1). We applied aperture photometry to Nova Sco 1994
and several nearby comparison stars  within the field of view. Johnson
$V$- and $R$-band magnitudes of these comparison stars were made
available to us by J.~Orosz. We determined the relative magnitude of
Nova Sco 1994 with respect to three stars having a range of  colours
[($V-R$)=0.49, 0.77 and 1.23]. Assuming that the colour correction
between the two filter systems is small ($<$0.05 mags; similar to the
accuracy of our photometry) and that Nova Sco 1994 has a  colour in
the same range as the comparison stars used, we estimate $R\sim$15.05
for Nova Sco 1994.

\begin{table}
\caption{Log of observations}
\begin{center}
\begin{tabular}{lccc}
Night $\#$ &  $\#$ of spectra & seeing ($\arcsec$) & Orbital phase  \\
           &                  &                    &  covered       \\
Night 1    &  11              & 2.0           &  0.722--0.834  \\  
Night 2    &  5               & 2.6           &  0.100--0.225  \\  
Night 3    &  11              & 2.3           &  0.501--0.592  \\  
Night 4    &  10              & 1.6           &  0.883--0.964  \\  
Night 5    &  10              & 3.0           &  0.270--0.353  \\
\end{tabular}
\end{center}
\end{table}
\begin{table}

\caption{H$\alpha$ equivalent width}
\begin{center}
\begin{tabular}{lcc}
Night $\#$ &  Orbital phase &   EW (\AA)      \\  
           &                &           \\  
Night 1    &     0.78       &   9.8$\pm$0.5   \\  
Night 2    &     0.16       & 11.9$\pm$0.3    \\  
Night 3    &     0.54       & 10.6$\pm$0.4    \\  
Night 4    &     0.93       &  3.8$\pm$010    \\  
Night 5    &     0.32       & 10.4$\pm$0.8    \\
\end{tabular}
\end{center}
\end{table}

\section{The spectra of Nova Sco 1994}

In Figure 1 we show the variance-weighted average and also the nightly
averages of the Nova Sco 1994 spectra. A strong H$\alpha$ emission
line (mean equivalent width of 7.5$\pm$0.06\AA) and a much
weaker He$\sc i$ 6678\AA\ (equivalent width of 0.4$\pm$0.03\AA) can
be seen. In Table 2 we list the H$\alpha$ equivalent width for
the nightly averages. The Fe$\sc i$ absorption blend at 6485, 6496,
6499, 6502\AA\ is also visible. These features are used to determine
the radial velocity of the secondary star (see section 4). The
6613\AA\ diffuse interstellar band is also present.

The emission lines in Nova Sco 1994 are double-peaked, which is
presumably a consequence of the system being at high inclination. We
can compare the observed peak-to-peak half separation of the H$\alpha$
emission line (which arises from the accretion disc) with the
projected velocity of the outer disc edge. In a binary system with a
mass ratio $>$ 0.25 it is generally assumed that the accretion disc
cannot grow larger than the tidal truncation radius, $r_d$ (Paczynski
1977; Whitehurst 1988; Osaki, Hirose \& Ichikawa 1993), which is
approximately given by $r_d = 0.60 a / (1+q)$ for 0.03 $< q <$ 1,
where $a$ is the binary separation (Warner 1995). Given the system
parameters ($P_{orb}$=2.62168 days; $q\sim 0.39$; $i\sim 69$ degrees;
$M_{1}\sim$6.7~$M_{\odot}$ see Shahbaz et al., 1999) the minimum value
for the projected velocity of the accretion disc rim is $\sim$394
\kms. The observed peak-to-peak half separation of the H$\alpha$
emission line (see Figure 1) in late June 1996 is 385$\pm8$ \kms
(measured by fitting the profile with a double Gaussian), which
implies that the accretion disc is close to its maximum possible size.
Soria et al. (1998) estimate the H$\alpha$ half peak-to-peak
separation to be $<$350/2 \kms and $<$550/2 \kms for their
August/September 1994 and June 1996 H$\alpha$ observations
respectively, velocities much lower than expected, suggesting that the
H$\alpha$ emission line arises from non-Keplerian regions/flows  
in the accretion disc.

\section{The radial velocity of the secondary star}

The radial velocities of the F-type secondary star in Nova Sco 1994
were measured from the spectra by the method of cross-correlation
(Tonry \& Davis 1979) with a template star. Prior to cross-correlation
the spectra were interpolated onto a logarithmic wavelength scale
(pixel size 55 \kms) using a $\sin\,x/x$ interpolation scheme to
minimize data smoothing (Stover et al. 1980), and then normalised.
The template star spectrum (HR2906; F6$\sc v$) was then artificially
broadened by 90 \kms (Shahbaz et al., 1999) to account for the
rotational velocity of the secondary star. Note that the orbital
smearing of the  Nova Sco 1994 spectra through the 1800s exposure is
at most only 10 \kms, much less than the resolution of the data.  Only
regions of the spectrum devoid of emission lines (6400-6520\AA) were
used in the cross-correlation.  The radial velocity of the template
star (derived using the position of the H$\alpha$ absorption line to
be $-$7 \kms) was then added to the radial velocities of Nova Sco 1994.

Using the orbital ephemeris given by van der Hooft et al (1998)  we
phase-folded and binned the heliocentric radial velocities (see Figure
2). From figure 2, it can be seen that the radial velocity measurement
at phase 0.2 does not fit the general pattern of the  sinusoidal
modulation present in the data. This data point was the total of three
radial velocity measurements taken on the second night 
(21st June 1996). Although the
seeing and quality of the spectra taken during this night were not as
good as the others, no obvious reason could be found as to why these
spectra gave  much lower radial velocities than expected.  A sine wave
fit to all the data points does not give an adequate fit
($\chi^{2}_{\nu}$=6.9).  However, removing the discrepant data point
and then  performing a sine wave fit yields a $\chi^{2}_{\nu}$ of 1.5,
a semi-amplitude $K_{2}=279 \pm 10$ \kms , systemic velocity
$\gamma=-155 \pm 7$ and a phase shift  of $-0.043 \pm 0.005 \phi$
(1-$\sigma$ errors are given). We also fitted the radial velocity curve
with an eccentric orbit, but found the fit to be less than 50 percent
significant.
 
\begin{table*}
\caption{Radial velocity and X-ray luminosity measurements }
\begin{center}
\begin{tabular}{cccc}
$K_{obs}$     &    $L_{x}$    & power-law index  & Corrected $L_{x}$ \\ 
(km~s$^{-1}$) & (erg~s$^{-1}$) & $N_h=0.89\times 10^{22}~cm^{-1}$ 
&  (erg~s$^{-1}$)    \\ 
               &                       & &    \\ 
215.5$\pm 2.4$ & $2.0\times 10^{32}$ ASCA (1-10 keV)     & 1.5  
& $2.1\times 10^{32}$ \\    
232$\pm 2$   &   $<2.4\times 10^{36}$ BATSE (20-350 kev) & 2.8   
& $<1.6\times 10^{37}$ \\ 
279$\pm 10$  & $6.8\times 10^{37}$ RXTE (2--12 keV)      & 2.8  
& $8.5\times 10^{37}$ \\
\end{tabular}
\end{center}
\end{table*}

\section{The effect of irradiation on the secondary star's radial velocity}

Three absorption line radial velocity curves have been obtained for
Nova Sco 1994, using the same absorption features of the  F6$\sc iv$
secondary star and the standard method of cross-correlation. However,
in each case the system was observed to be in a different X-ray state.
A sinusoidal fit to the outburst data taken in April/May 1995 of Orosz
\& Bailyn (1998) gives a radial velocity semi-amplitude  of
$K_{obs}$=230$\pm 2$ \kms. During this period BATSE did not detect the
source, so we can only put an upper limit of 2.4$\times$10$^{36}$
\ergs ($<$0.03~photons~erg~cm$^{-2}$~s$^{-1}$ in the BATSE 20-350 keV
energy range) to the X-ray luminosity of the source.  This upper limit
alone does not allow us to state unequivocally that the source was not
active at X-ray energies, but optical observations suggest  that the
source was not in quiescence ($V$=16.5; Orosz \& Bailyn 1998). In
section 4 we determined $K_2$=279$\pm10$\kms from data taken in June
1996 when RXTE ASM (2-12 keV) observations give an X-ray luminosity of
$L_x$=6.8$\times$10$^{37}$ \ergs, and the $R$-band brightness was
$\sim$ 1 mag brighter than its quiescent value.  The BATSE (20--350
keV) count rate was at least a factor of 4 higher than in April/May
1995. Shahbaz et al. (1999) determined the true radial velocity of
the secondary star ($K_2$=215.5$\pm2.4$ \kms) in 1998 May/June,
when the source was finally in optical quiescence. The only X-ray
quiescent  observations were obtained during March 1996 using ASCA
(1-10 keV; Robinson et al., 1997) which gave $L_x$=2$\times$10$^{32}$
\ergs.

In Figure 3 we show the observed radial velocity amplitudes relative
to the quiescent value as a function of the observed  X-ray luminosity
at the time of the measurements.  We have converted the X-ray
luminosities, which were observed with different instruments, into a
common energy range (0.4--10 keV) using a hydrogen column density of
$N_h=0.89\times 10^{22}$ cm$^{-2}$ and a photon power-law model with
indices 2.8 and 1.5 for the X-ray high and quiescent states
respectively  (see Table 3;  Zhang et al., 1997; Robinson et al.,
1997; Hameury et al., 1997).  This energy range is where we expect the
total radiated power for X-ray transients in both outburst and
quiescence to lie (Chen, Shrader \& Livio 1997).  Note that there is a
correlation between X-ray luminosity and the observed radial velocity
semi-amplitude; the higher the X-ray luminosity the larger the
observed radial velocity semi-amplitude, exactly as expected.  We can
use our model to estimate the X-ray luminosity at the  time when Orosz
\& Bailyn (1997) took their radial velocity measurements.  We find
$L_{x} \sim 5\times 10^{35}$ \ergs which is consistent with the BATSE
upper limit.

\section{Irradiation of the secondary star}

It has been known for some time, especially in studies of dwarf novae
and polars, that substantial heating of the secondary star shifts the
effective centre of the secondary, weighted by the strength of the
absorption lines, from the centre of mass of the star. This results in
a significant distortion of the radial velocity curve leading to a
spuriously high semi-amplitude and a radial velocity curve that may be
eccentric.  Davey \& Smith (1992) describe a procedure for detecting
the effects of irradiation on the radial velocity curve of the
secondary star, whereby one tests the significance of an eccentricity
in the orbital solution.  However, it should be noted that, although
our data does not allow this eccentricity test, due to the poor 
orbital phase coverage, we can use the
spuriously high radial velocity semi-amplitude to show that X-ray
heating is present.

In order to investigate the effects of X-ray heating on the  secondary
star's radial velocity curve we used the model described by Phillips,
Shahbaz \& Podsiadlowski (1999).  The model uses a crude treatment for
X-ray heating, since no satisfactory robust model exists for the effects of
external heating in stars. However, it serves to illustrate the extreme 
effects of X-ray heating.
It should be noted that the first order model of X-ray heating by
Brett \& Smith (1993), which does not include energy transport effects, 
does show that the whole temperature structure of the outer layers of the 
secondary is upset by external heating.
Figure 3 shows the effects of
different amounts of X-ray luminosity on the secondary star's radial
velocity amplitude. $K_{\rm obs}$ is computed by fitting the predicted
curve with an eccentric orbit.   The regions on the secondary star
that are heated  do not contribute to the absorption line flux.  The
maximum possible change that irradiation can have on $K_{obs}$, based
purely on geometry, is 15 per cent.  However, from our data presented
in this paper, we observe $\Delta K_2/K_2$=0.30$\pm$0.05, which when
compared with maximum possible value based on geometry, is significant
at the 3-$\sigma$ level.

In Figure 4 we show how much of the secondary star's surface needs to
be heated in order to produce the observed radial velocity amplitude.
We find that based purely on geometry 35 per cent of the secondary
star's surface is directly heated by X-rays produced at the compact
object. (This fraction only depends on the shape and size of the 
secondary star, which in turn is determind by the $q$. Using the extreme
values for $q$ (Shahhaz et al., 1999), we find that this 
fraction changes by less than 1 per cent.)
However, in order to produce the observed large radial
velocity semi-amplitude, 60$^{+9}_{-7}$ per cent of the 
secondary star needs to be heated.  
The 1-$\sigma$ uncertanties quoted here were estinated using the 
1-$\sigma$ uncertanties in $\Delta K_2/K_2$.
This result may
seem surprising at first, since one expects only the regions  of the
secondary star facing the compact object to be irradiated and yet our
result implies that some of the regions not directly seen by the
compact object are also affected by irradiation. However, one should
note that effects such as X-ray scattering and irradiation-induced 
flows on the surface
of the secondary star (Phillips \& Podsiadlowski 1999) can increase
the fraction of the secondary star that responds to the X-ray
source. Note that the regions on the secondary star that are shadowed
by the accretion disc will be indirectly heated by such mechanisms.
Therefore $K_{\rm obs}$ can be larger than that expected from heating
the inner face of the secondary star alone.

\section{Discussion}

The existence of circulation in rotating stars  
was first proved in 1924 by von Zeipel (von Zeipel 1924).
He demonstrated that
for a rotating homogeneous star, the radiative transport equation and
equation of conservation of energy cannot be fulfilled
simultaneously. This results in the formation of meridional motions.
In order to maintain a stationary state as assumed, one has to demand
that these meridional motions contribute to the energy transport. In
the case of an irradiated rotating star, the situation is still more
complicated, since the radiation will induce additional circulation
currents.

Evidence for the existence of significant irradiation-driven
circulation is provided by several sources. For example, the analysis
of the optical light curve of HZ Herculis has shown this to be heated
by its accompanying X-ray source HER X-1. Although the main features
of the optical light variation are well understood (HZ Her is bright
when the X-ray source is in front of it, its brightness is reduced
during the occultation of the X-ray source by the secondary), the
minimum at phase 0.0 is sharper than expected and indicates some
additional source of optical radiation at this phase. 
Strittmatter et al. (1973)
tried to explain this via the illumination of the disk by HZ
Her. Other attempts were made by Pringle (1973) and
Bahcall, Joss \& Avni (1974). 
However, the most successful explanation was due to
Kippenhahn \& Thomas (1979).
They estimated the energy transported from the X-ray
illuminated part of the stellar surface to the shadowed side, and
demonstrated that the minimum at phase 0.0 could be reasonably well
accounted for (X-ray heating without horizontal transport leads to a
flat minimum at phase 0.0).

In addition, Schandl et al. (1997) found circulation to be necessary
in order to accurately model the optical light curve of CAL 87, an
eclipsing supersoft X-ray source. They calculated the light curve based
on the assumption that an accreting, steadily burning white dwarf
irradiates the accretion disk and the secondary star, as suggested by
van den Heuvel et al. (1992). A simple description of energy
transport on the secondary surface was used 
and then integrated over the whole
surface, while conserving the total luminosity.  They found that
significant energy transport of the irradiated flux to non-illuminated
parts on the secondary surface is required to simulate the observed
lightcurve, particularly around the primary eclipse, when the
shadowed hemisphere of the secondary is in view. 

Recent models for irradiation-induced flows in binary stars 
have been computed by Martin \& Davey (1995). They considered circulation in 
gently-heated secondary stars (where the incident flux is less 
than the intrinsic flux). Their 2-dimensional calculations 
included the effects of the Coriolis force and showed upwelling of 
hot material being carried preferentially towards the direction of rotation of
the star.
They also concluded that all secondary stars should show asymmetric 
heating, because of the presence of Coriolis forces.
Phillips (1999) has recently extended the study of circulation 
to 3-dimensions. As well as including the effects of X-ray irradiation i.e.
the anisotropic heating of the irradiated surface, and the effects of
surface radiation stress, he also considers the 
large-scale effects of the rotation of the system and includes 
an approximate treatment of the Coriolis force. His results suggest a 
realistic analysis of the Coriolis force is essential 
for a full description of stellar circulation. 

In order to study the extent of irradiation of the secondary star one
requires good quality spectrophotometric studies throughout an X-ray
outburst, during which the level of X-ray irradiation and induced heating 
changes. This will allow the surface intensity distribution across
the secondary star to be mapped (see Rutten \& Dhillon 1994 and Davey
\& Smith 1996), from which effects such as irradiation-induced
circulation or star-spots can be investigated.

\section*{Acknowledgements}

We would like to thank the referee, Prof. Robert Smith for useful comments.

\begin{figure*}
\rotate[l]{\epsfxsize=500pt \epsfbox[00 00 700 750]{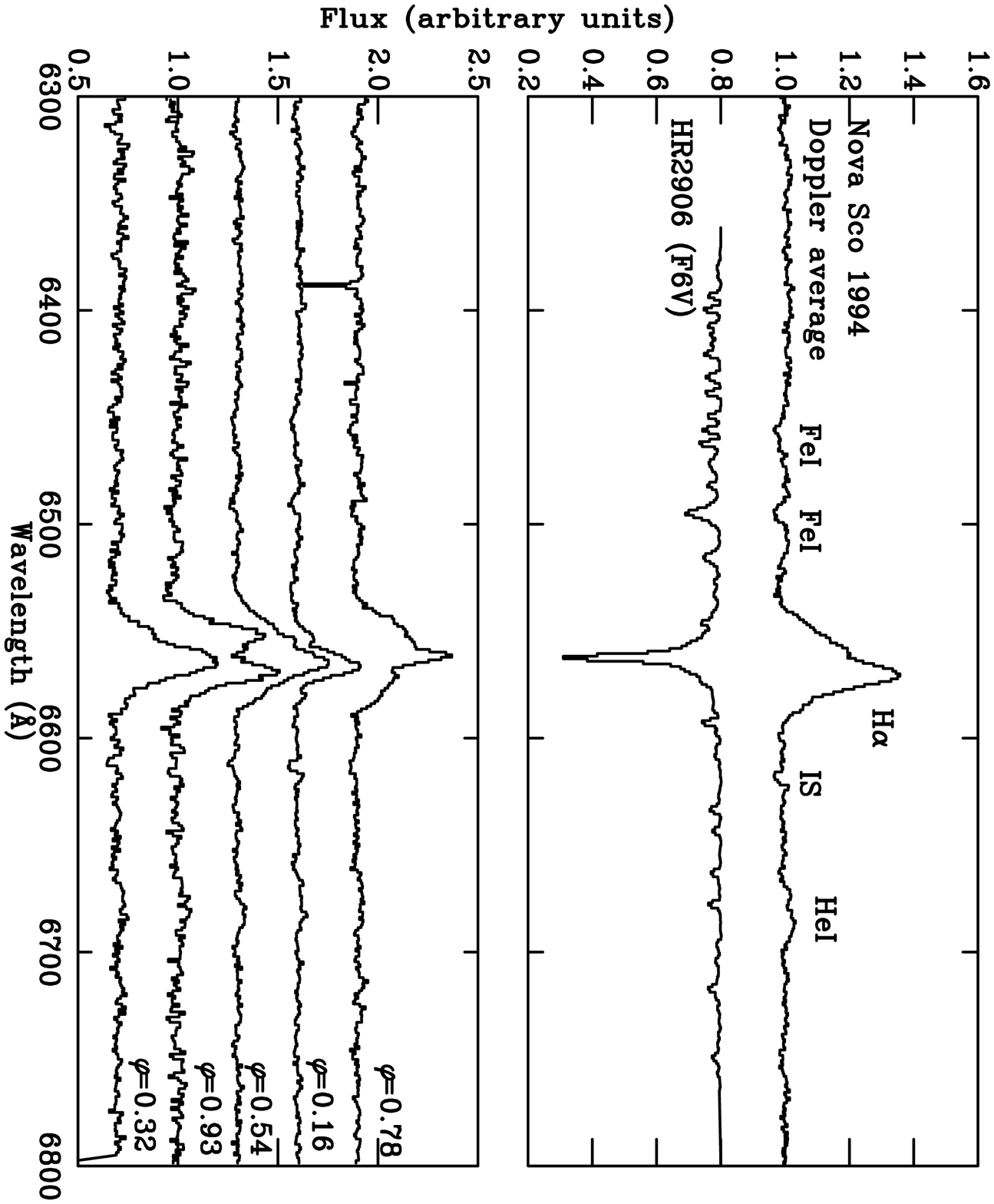}}
\caption{Top: Doppler average spectrum of Nova Sco 1994 in the rest frame of
the secondary star and an 
F6 star (HR2906).  Bottom:
Nightly averaged spectra of Nova Sco 1994 (first night at the top,
last night at the bottom).  The spectra have been normalized and shifted
vertically for clarity. IS indicates the interstellar 6613\AA\ line.}
\end{figure*}

\begin{figure*}
\rotate[l]{\epsfxsize=500pt \epsfbox[00 00 700 750]{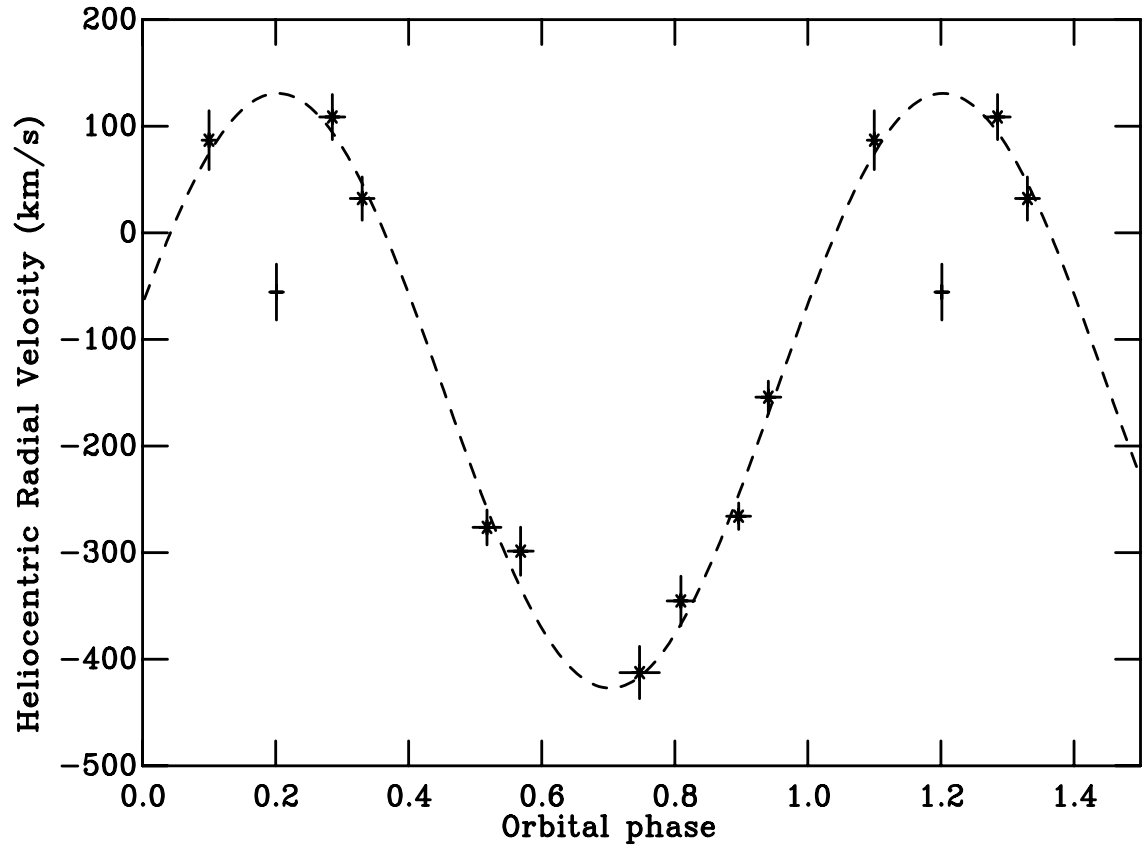}}
\caption{ The radial velocity curve  of the secondary star in Nova Sco
1994. The dashed curve is a sinusoidal fit to the data points marked
with a star.  The data have been folded on the orbital ephemeris given
by van der Hooft et al., (1998) and 1.5 orbital cycles are shown.}
\end{figure*}

\begin{figure*}
\rotate[l]{\epsfxsize=500pt \epsfbox[00 00 700 750]{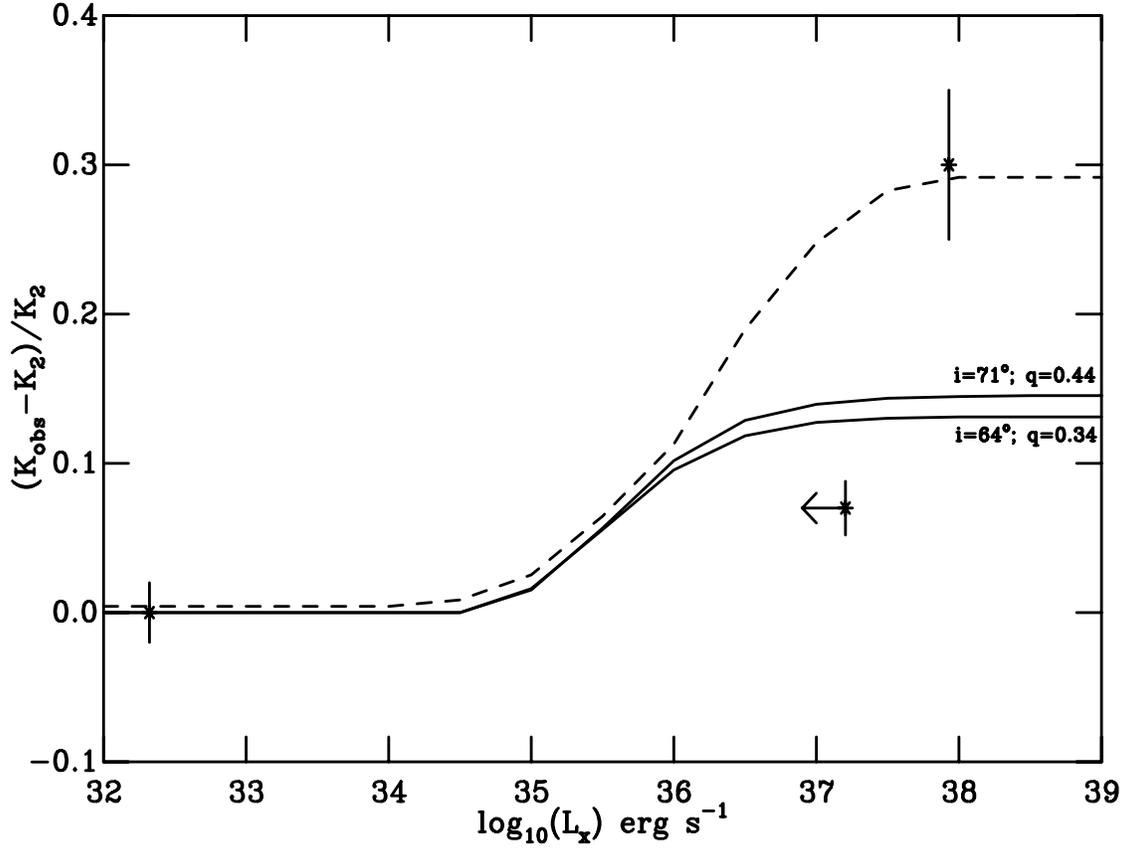}}
\caption{ The effects of different amounts of X-ray heating on the
secondary star's radial velocity semi-amplitude. We show the
fractional change in $K_{\rm obs}$ as a function of $L_{x}$. $K_{\rm
obs}$ is computed by fitting the  predicted curve with an eccentric
orbit. The three observed radial velocity measurements are also shown.
The two lower curves show the effects of X-ray heating  based purely
on geometry i.e. only those elements of area on the secondary star
that are directly seen by the X-ray source are irradiated.   The top
curve (dashed line) show the effects of indirect X-ray heating,
calculated by extending the radiation horizon as seen by the X-ray
source by a further 24 degrees (see Figure 4). The effect of this
additional heating is to produce a value for $K_{\rm obs}$ which is 30
per cent larger than expected   purely based on geometry.}

\end{figure*}

\begin{figure*}
\rotate[r]{\epsfxsize=500pt \epsfbox[00 00 700 750]{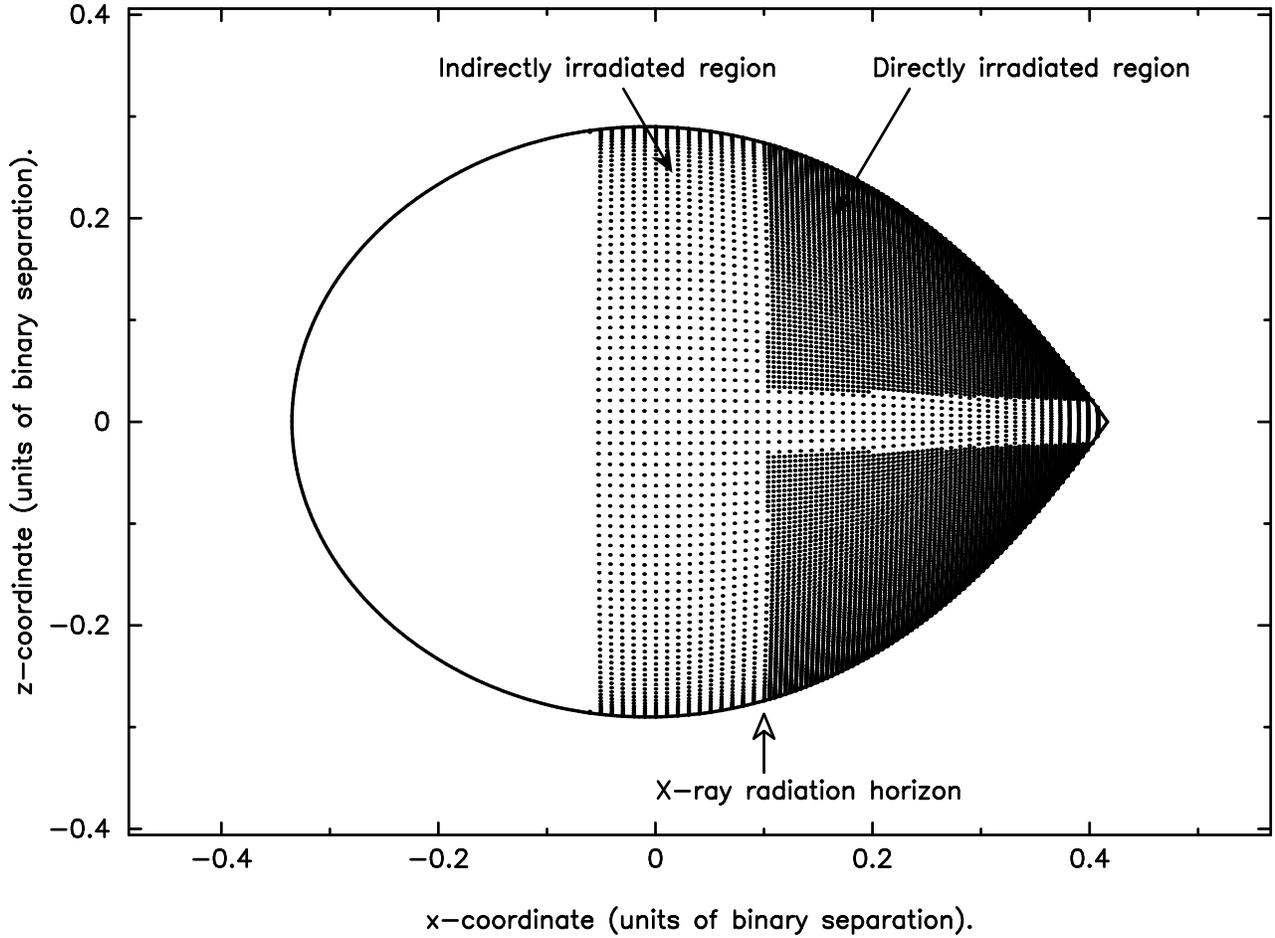}}
\caption{The irradiated secondary star's Roche lobe in the ($x-z$)
plane. The compact object is at coordinates (1,0). We have assumed
$L_x=8.5\times 10^{37}$ erg~s$^{-1}$ and the extreme geometrical case
with a mass ratio of $q$=0.44, an inclination angle of $i$=71$^\circ$
and a disc angle of 2$^\circ$. The dense shaded region shows the area
(35 per cent) that is irradiated directly  by X-rays produced at the
compact object; these regions do not contribute to the observed
absorption line flux. The less dense region shows the area (60 per
cent) which must also be heated indirectly in order to produce the
large observed radial velocity semi-amplitude. }
\end{figure*}

\end{document}